# Optical control of internal electric fields in band-gap graded InGaN nanowires

N. Erhard[1,2], A.T.M. Golam Sarwar[3,4], F. Yang[4], D. W. McComb[4], R. C. Myers[3,4], and A. W. Holleitner[1,2]

[1] Walter Schottky Institut and Physik-Department, Technische Universität München, D-85748 Garching, Germany
[2] Nanosystems Initiative Munich (NIM), Schellingstr. 4, 80799 München, Germany
[3] Department of Electrical and Computer Engineering, The Ohio State University, Columbus, OH, 43210, USA
[4] Department of Materials Science and Engineering, The Ohio State University, Columbus, OH, 43210, USA

**InGaN nanowires are suitable building blocks for many future optoelectronic devices. We show that a linear grading of the indium content along the nanowire axis from GaN to InN introduces an internal electric field evoking a photocurrent. Consistent with quantitative band structure simulations we observe a sign change in the measured photocurrent as a function of photon flux. This negative differential photocurrent opens the path to a new type of nanowire-based photodetector. We demonstrate that the photocurrent response of the nanowires is as fast as 1.5 ps.**

In recent years, $In_xGa_{1-x}N$ nanowires have attracted increasing attention as building blocks for optoelectronic devices such as solar cells[1,2], LEDs[3,4], white light emitting LEDs[5,6], lasers[7], photodetectors[8] and optochemical gas sensors[9] making $In_xGa_{1-x}N$ nanowires ideal candidates for integrated photonic circuits[10]. The reason why $In_xGa_{1-x}N$ nanowires are so well suited for optoelectronic devices, is the fact that $In_xGa_{1-x}N$ is a semiconductor alloy with a direct band gap[11] tunable from 0.7 eV (InN)[12,13] to 3.4 eV (GaN)[14] covering the whole visible spectrum from the near infrared to the near ultraviolet. Furthermore, the nanowire geometry allows a strain-relaxed growth. This suppresses the tendency towards phase separation which is problematic in thin film growth.[15,16] A full tunability of the In content in $In_xGa_{1-x}N$ is difficult to achieve in thin film growth due to a lack of lattice-matched native substrates[17] and the large lattice mismatch of 11% between GaN and InN.[18] In contrast single-crystal $In_xGa_{1-x}N$ nanowires across the entire compositional range from $x$ = 0 to 1 can be grown[19].

It has been shown that 1 µm long N-face[21] InGaN-nanowires with an approximately linearly graded In content along the nanowire axis covering the entire compositional range can be successfully grown by catalyst-free plasma-assisted molecular beam epitaxy.[22] Varying the In content within a single $In_xGa_{1-x}N$ nanowire from $x$ = 0 to 1 enables optoelectronic devices with an active region with a gradually tunable, direct band gap desirable for a good conversion efficiency in solar cells. In addition, the graded In composition gives rise to intrinsic electric fields in the nanowire.[23] These are favorable for applications in solar cells and sensitive photodetectors. Generally, the recombination lifetimes in GaN/InGaN nanowires are in the nanosecond regime.[6] Recent publications report a photo response only in the millisecond regime and a persistent photocurrent in GaN/InGaN-nanowire devices.[8] Both time scales are too slow for fast photodetectors operating in the GHz to THz regime (ns to ps).

Here we demonstrate a new type of ultrafast semiconductor photodetector based on the built in electric field caused by a band gap gradient. This gradient is induced by a linear increase of the In content along the InGaN nanowire axis. The relative electric field contributions appearing in the conduction and valence band are tuned in sign and magnitude by applying an optical pulse. This optical gating manifests as an observed sign change in the photocurrent as a function of optical power density. Moreover, the linearly graded InGaN nanowires exhibit an ultrafast photocurrent response of 1.5 ps making them good candidates for fast photodetectors. Our results demonstrate a new scheme for optical gating of nanowire devices with graded band gaps, namely, the intensity of an optical pulse can be used to alter the sign of the transient photocurrent by switching the sign of the built in electric field, an effect that we name 'negative differential photocurrent' due to its functional similarity to negative differential resistance observed in tunneling diodes.

To begin, 1 µm long graded InGaN nanowires with a diameter of about 50 nm to 100 nm are grown by plasma-assisted molecular beam epitaxy on a Si(111) substrate (Figure 1a and cf. Supporting Information).[22] The sample for scanning transmission electron microscopy (STEM) was prepared with a Helios NanoLab 600 dual-beam focused ion beam (FIB). STEM images and X-ray energy dispersive spectroscopy (EDS) were carried out on an FEI image corrected Titan3™ G2 60-300 S/TEM equipped with a quad silicon drift detector operated at 300kV. The continuous grading of the In content is verified by performing STEM-EDS (Figure 1b). The quantitative elemental maps show that the nanowire base is nearly 100% GaN. The In content is increased along the nanowire axis resulting in a dominating InN phase at the nanowire tip (Figure 1c). For the optoelectronic measurements, the InGaN nanowires are mechanically transferred onto a pre-processed sapphire substrate. A single InGaN nanowire is electrically contacted by Pt contacts fabricated by electron beam induced deposition (EBID) utilizing [$(CH_3C_5H_4)*(CH_3)_3Pt$] as a precursor in an NVision 40 (Inset in Figure 2a). The presented optoelectronic measurements are performed in a cryostat at 77K at about $10^{-6}$ mbar. The dark current-voltage characteristic is shown in Figure 2a. The optical excitation occurs by focusing a mode-locked laser with an energy of $E_{photon}$ = 3.1 eV, a pulse duration of 160 fs, and a repetition frequency of 76 MHz with the help of a microscope objective onto the nanowire. The photocurrent $I_{Photo}$ is measured by a lock-in technique (see Methods).

Due to the mechanical transfer onto the sapphire substrate the investigated nanowire fragment is about 400 nm long (Inset Figure 2a). From wavelength dependent photocurrent measurements, the smallest band gap is estimated to be 1.13 eV corresponding to a maximum In content of about $x$ = 0.74 ± 0.2 [24] (cf. supporting information). Figure 2b shows the calculated equilibrium band structure of the investigated nanowire device, with a linear In content increase from $x$ = 0.3 to 0.74 from left to right assuming a linearly graded In$_x$Ga$_{1-x}$N-nanowire (cf. supporting information). The intrinsic InGaN nanowire is assumed to be n-type because of the expected high background doping at low growth temperatures (375°C). Moreover, for In rich In$_x$Ga$_{1-x}$N with $x$ > 0.49, the surface depletion layer fades to a surface accumulation layer.[25] In the metal-semiconductor-metal geometry, the equilibrium band structure of the semiconducting nanowire exhibits Schottky barriers at the Pt/nanowire contacts. The Schottky barrier for the Ga rich end of the nanowire is larger than for the In rich end. Far from the Schottky barriers, the conduction band bottom edge $E_C$ remains flat and parallel to the Fermi energy level $E_F$ due to the n-type nature of the nanowire. The difference in the bandgap caused by the graded In content results in an up-bending of the valence band top edge $E_V$ towards the In rich end of the nanowire. This band profile generates an electric field of about $3 \cdot 10^4$ V/cm for the charge carriers in the valence band.

The calculated equilibrium band structure is consistent with the dark current measurements performed on the InGaN nanowire device (Figure 2a). The dark current-voltage characteristic exhibits an s-shape as expected for a metal-semiconductor-metal device with a semiconductor nanowire.[26] For low bias, the applied voltage drops across the reverse biased Schottky barrier, resulting in a small dark current due to tunneling. The reverse tunneling current is not negligible and for higher bias, the current is dominated by the linear resistance of the nanowire.[26] In the dark current measurement, the influence of both Schottky barriers is comparable in contradiction to the simulated equilibrium band structure. This can be explained by the further influence of surface Fermi level pinning which is not considered in the simulations. In particular, donor like point defects, e.g. introduced during growth or by the electron radiation during the EBID, can lead to such a Fermi level pinning [27].

The dark current is an electron current due to the *n*-type nature of the InGaN nanowire. Thus, the band bending in the conduction band ($E_V$) does not need to be regarded here. In contrast, for photocurrent measurements, both electron and hole currents need to be considered, as the density of photo-generated electrons and holes is equal ($n_\text{Photo} = p_\text{Photo}$). In turn, we can write

$$I_\text{Photo} = en_\text{Photo}(\mu_\text{electron}\mathcal{E}_\text{electron} - \mu_\text{hole}\mathcal{E}_\text{hole}), \quad (1)$$

with $e$ the elementary charge, $\mu_\text{electron}$ ($\mu_\text{hole}$) the mobility of electrons (holes) and $\mathcal{E}_\text{electron}$ ($\mathcal{E}_\text{hole}$) the quasi electric field in the conduction (valence) band edge. Figure 2c depicts the spatially resolved photocurrent of the investigated InGaN nanowire for an applied bias of 5 V. The longitudinal shape of the photocurrent is due to the elliptical spot of the laser. All further photocurrent measurements are performed for excitation at the spatial maximum of $I_\text{Photo}$. Thus, we ensure that the entire nanowire is excited by the laser.

When we measure the photocurrent vs. the applied bias for an increasing photon density, we observe a shift of the zero photocurrent-intersection ($I_\text{Photo}$ = 0 A) towards positive bias for small laser power densities (white dots in Figure 3a). For 0.9 MW/cm², this intersection occurs at about 3 V which corresponds to an intrinsic electric field of about $7 \cdot 10^4$ V/cm present in the unbiased nanowire device at such a power density. We note that the internal electric field in these graded nanowires is more than one order of magnitude lower than what is usually observed in GaN/AlGaN[28] or GaN/InGaN[29] abrupt heterostructures. However, the extracted field value is larger than expected from the gradient in $E_V$ of the equilibrium band structure ($3 \cdot 10^4$ V/cm) in Figure 2a. We explain this by the fact that the simulated band structure does not consider Fermi level pinning at the EBID Schottky contacts. For an increasing laser power density, the zero photocurrent-intersection approaches a negative value close to 0 V (Figure 3b). From these data, we can conclude that the steady-state energy band diagram is strongly changed by the incident light, and that this change depends on the laser power.

We now focus on the power density dependent photocurrent at zero bias (Figure 4a). For small power densities (< 4 MW/cm²) the photocurrent is negative, reaching its most negative value at about 2 MW/cm². By further increasing the power density, the photocurrent reduces to zero and it becomes positive at about 4 MW/cm². The corresponding simulated band structures for the nanowire under 160 fs pulsed excitation are shown in Figure 4b-d. The simulations reveal the following scenario.

At low power densities (< 2 MW/cm²), the photocurrent for zero bias is negative because of a dominating hole current to the right contact which is the In rich end of the nanowire. This observation is in accordance to with the equilibrium band structure in Figure 2b, because a major part of the photogenerated charge carriers is generated in the main fraction of the nanowire, where $E_C$ is flat and $E_V$ bends upwards towards the In rich part of the nanowire resulting in a hole-photocurrent to the right. In addition, at the Schottky contacts, photocurrents are generated which counteract (left contact) or enhance (right contact) the hole-photocurrent. But as the Schottky barriers cover only the smaller part of the nanowire area, the number of charge carriers photogenerated inside the Schottky barrier is smaller than the number generated in the main area. Thus, the hole-photocurrent to the In rich end of the nanowire dominates at low power densities.

At a power density of 2 MW/cm$^2$ (Figure 4b), the presence of excess charge carriers (photogenerated electrons and holes) lower the Schottky barriers considerably compared to the equilibrium band structure (Figure 2b). In particular, the Schottky barrier on the Ga rich end of the nanowire (left contact) is reduced, resulting in an increased hole photocurrent as the hole barrier on the left contact is lowered (Figure 4b). The total photocurrent reaches its most negative value.

When the power density is further increased toward 4 MW/cm$^2$ (Figure 4c) or even 10 MW/cm$^2$ (Figure 4d), the Schottky barriers disappear and the nanowire band structure is noticeably altered. This can be explained by the fact that for nitride material systems, the electron mobility and thereby the electron diffusivity is far higher than that for holes.[30,31] Therefore, the photogenerated holes stay longer in the nanowire than the photogenerated electrons before they recombine or arrive at the In rich contact. This non-equilibrium accumulation of photogenerated holes leads to a screening of the internal electric field in the valence band. Since the band gap gradient caused by the graded In content must remain, the internal electric field shifts into the conduction band. This field favors drift of electrons to the In rich right contact. At the same time a maximum in $E_V$ arises which traps the photogenerated holes inside the nanowire. In total, the current of photogenerated holes decreases and the current of photogenerated electrons to the right contact increases with power density. At 4 MW/cm$^2$ (Figure 4c and arrow in Figure 4a), this electron current exceeds the hole current and the photocurrent becomes positive for 0 V applied bias. This negative differential photocurrent behavior clearly explains the sign change of the measured $I_{\text{Photo}}$ shown in Figure 4a.

With these insights, we discuss the bias dependence of Figure 3 in more detail. To this end, we consider the bias dependent photocurrent for two power densities (Figure 5a). We observe a change in the shape of the $I_{\text{Photo}}$-$V_{\text{sd}}$- characteristic. For low power densities (1 MW/cm$^2$), the characteristic has a linear shape within the signal-to-noise ratio. The zero-photocurrent intersection is positive (arrow in Figure 5a). For 4 MW/cm$^2$, the zero-photocurrent intersection becomes almost zero, and it changes sign for larger power intensities (compare Figures 3b and 5a). In addition, an s-shape becomes distinguishable with a large slope for low bias ($|V_{\text{SD}}|$ < 0.2 V in Figure 5a) and a smaller slope for larger bias. The dependence for low bias can be explained by the fact, that the bias increases the electric field inside the nanowire. The higher this overall field the more photogenerated charge carriers, in particular holes, can propagate to the contacts before the recombination takes place. This effect saturates for higher voltages as at some point all photogenerated charge carriers contribute to the photocurrent.[32,33] Thus, the photocurrent $I_{\text{Photo}}$ (equation 1) is only further increased by the electric field induced by the applied bias voltage. To further describe the process, we extract an off-set photocurrent $\Delta I$ (Figure 5a). The value $\Delta I$ is a measure of the number of photogenerated charge carriers as it marks the turning point where all photogenerated charge carriers contribute to the photocurrent. Consistent with this interpretation, $\Delta I$ saturates with increasing power density (Figure 5b) which means that also the number of photogenerated charge carriers saturates. The saturation can be explained by a bleaching effect by state filling at the band edges.[34–36]

Although we have not explicitly included the effect of In-rich clusters in our energy band modeling, it is speculated that such clusters would create local potential minima that capture photo-generated carriers and enhance the recombination. As a result, if the number of In-rich clusters were increased, then the minority carrier lifetime for both electrons and holes would be expected to decrease, lowering the photocurrent. The presence of clusters could account for our observation of a lower photo-current than what is predicted using a cluster-free model. Although clusters affect the local potential and act as recombination sites, they should not affect the key point of the proposed model, that the average internal electric field is tuned in sign and magnitude with the excitation intensity. A rigorous study is needed to examine the formation of In-rich clusters in these nanowires and its effect on photo-current, which is beyond the scope of the present study.

So far, the timescales of the photocurrent response in InGaN nanowires are reported to be in the second to millisecond regime.[8] From time-resolved photocurrent measurements on GaAs-[37] and InAs-nanowires[38], it is known that electric fields, also present in our InGaN nanowires, can generate an ultrafast displacement current in the picosecond regime[39]. We perform ps-time-resolved photocurrent measurements, as described previously (supporting information)[37,38] to verify such an ultrafast photocurrent in the graded InGaN nanowire. Figure 6 shows that the InGaN nanowire indeed exhibits a photocurrent response with a full width of half maximum (FWHM) of about 1.5 ps consistent with the simulated photocurrent response (red line). The amplitude of the simulation is normalized to the measured data as the absolute photocurrent value is modified by the readout circuit. The simulation exhibits fast oscillations at 0 ps which are simulation artefacts due to the laser excitation pulse in the first 160 fs. These fast fs-oscillations are beyond our resolution limit which is ~1 ps. The observed photocurrent is consistent with a displacement current due to electric fields as discussed in the context of Figures 3, 4 and 5. The time-resolved photocurrent measurements are performed at a high laser power density of 115 MW/cm² to compensate the low signal to noise ratio. In this regime, the photocurrent stems mainly from the photogenerated electrons and the displacement of the electric field in the conduction band.

In summary, we demonstrate a new type of semiconductor photodetector based on InGaN nanowires axially graded from GaN to InN exhibiting an internal photo-tunable quasi-electric field giving rise to a negative differential photocurrent. The results are corroborated by numerical simulations of the energy band structure. By increasing the laser power density, the band structure is tilted by the presence of photogenerated holes. This leads to a sign change of the photocurrent of the unbiased nanowire by increasing the laser power. The InGaN nanowires exhibit an ultrafast photocurrent with a FWHM of about 1.5 ps. Therefore the investigated InGaN nanowires are interesting for a new generation of ultrafast nanowire-based photodetectors.

We gratefully acknowledge financial support of the ERC-grant NanoREAL and the Center for NanoScience (CeNS). A.G.S. and R.C.M. were supported by the National Science Foundation CAREER award (DMR-1055164).

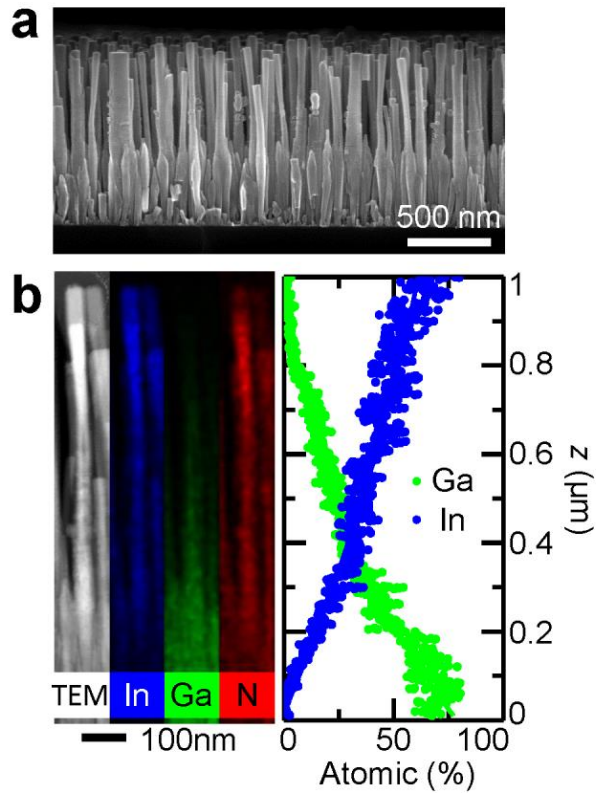

**Figure 1.** (a) Scanning electron microscope (SEM) image of InGaN nanowires grown on a Si (111) wafer. (b) Scanning transmission electron microscope (STEM) image (left) of the as-grown InGaN nanowires and the corresponding quantitative elemental maps for In (blue), Ga (Green) and N (red) extracted from the STEM-EDS spectrum image. (c) Concentration profile showing the Ga- and In-content along the nanowire axis from the base (0 µm) to the tip (1 µm).

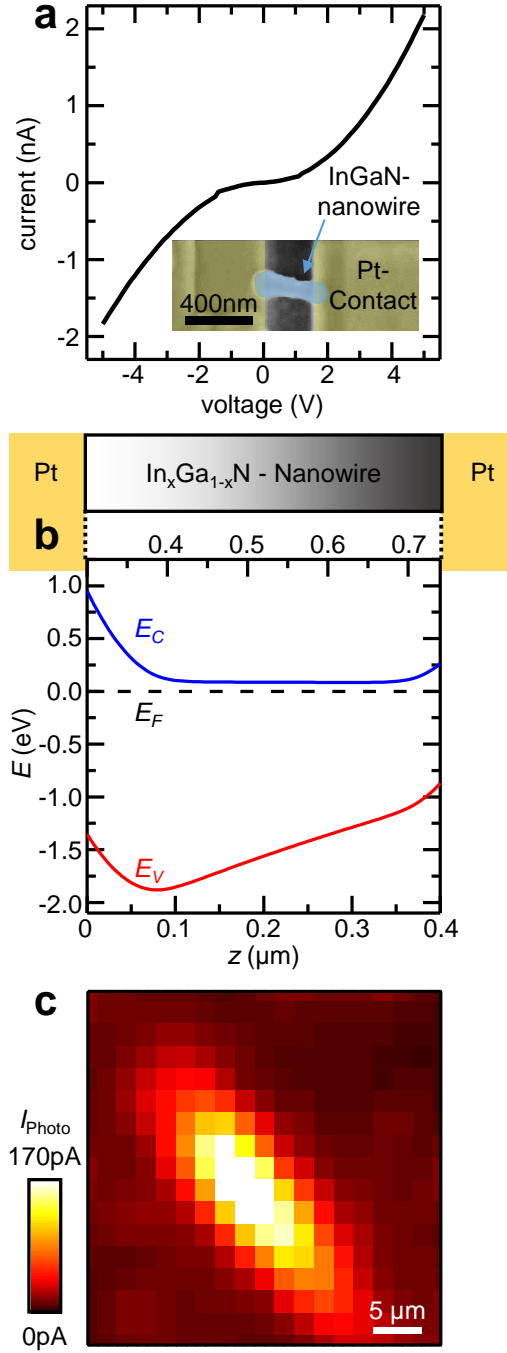

**Figure 2**. (a) Dark current voltage characteristic of the InGaN nanowire at 77 K. Inset shows a false color SEM image of the investigated nanowire contacted by EBID. (b) Device geometry for optoelectronic measurements (top) and calculated equilibrium band structure of the investigated $In_xGa_{1-x}N$ nanowire (bottom) with a linear graded In content from x = 0.30 (left contact) to 0.74 (right contact). $E_C$ ($E_V$) is the conduction bottom (valence top) band edge depicted in blue (red). The Fermi level $E_F$ is represented by the dashed black line. (c) Spatially resolved photocurrent of the InGaN nanowire for $V_{SD}$ = 5 V at 77 K illuminated by a longitudinal shaped laser spot ($E_{photon}$ = 3.1 eV, $p$ = 58 MW/cm$^2$).

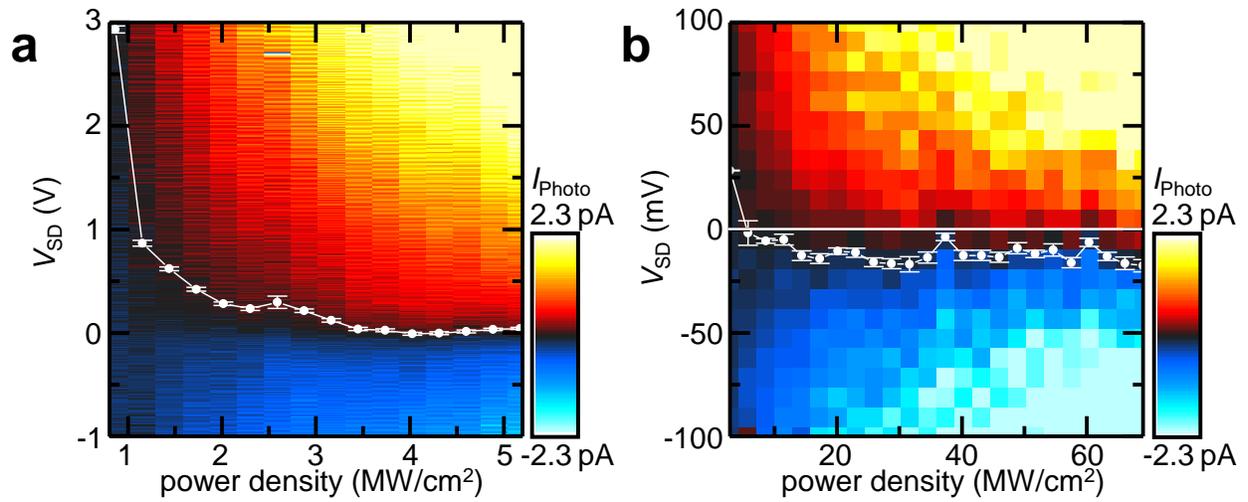

**Figure 3.** Time-averaged photocurrent $I_{Photo}$ versus power density and voltage $V_{SD}$ for (a) low and (b) high laser power densities. All power densities are calculated for a temporal Gaussian laser pulse with a FWHM of 160 fs. White line marks the level where $I_{Photo}$ = 0 A ($E_{photon}$ = 3.1 eV, $T_{bath}$ = 77 K).

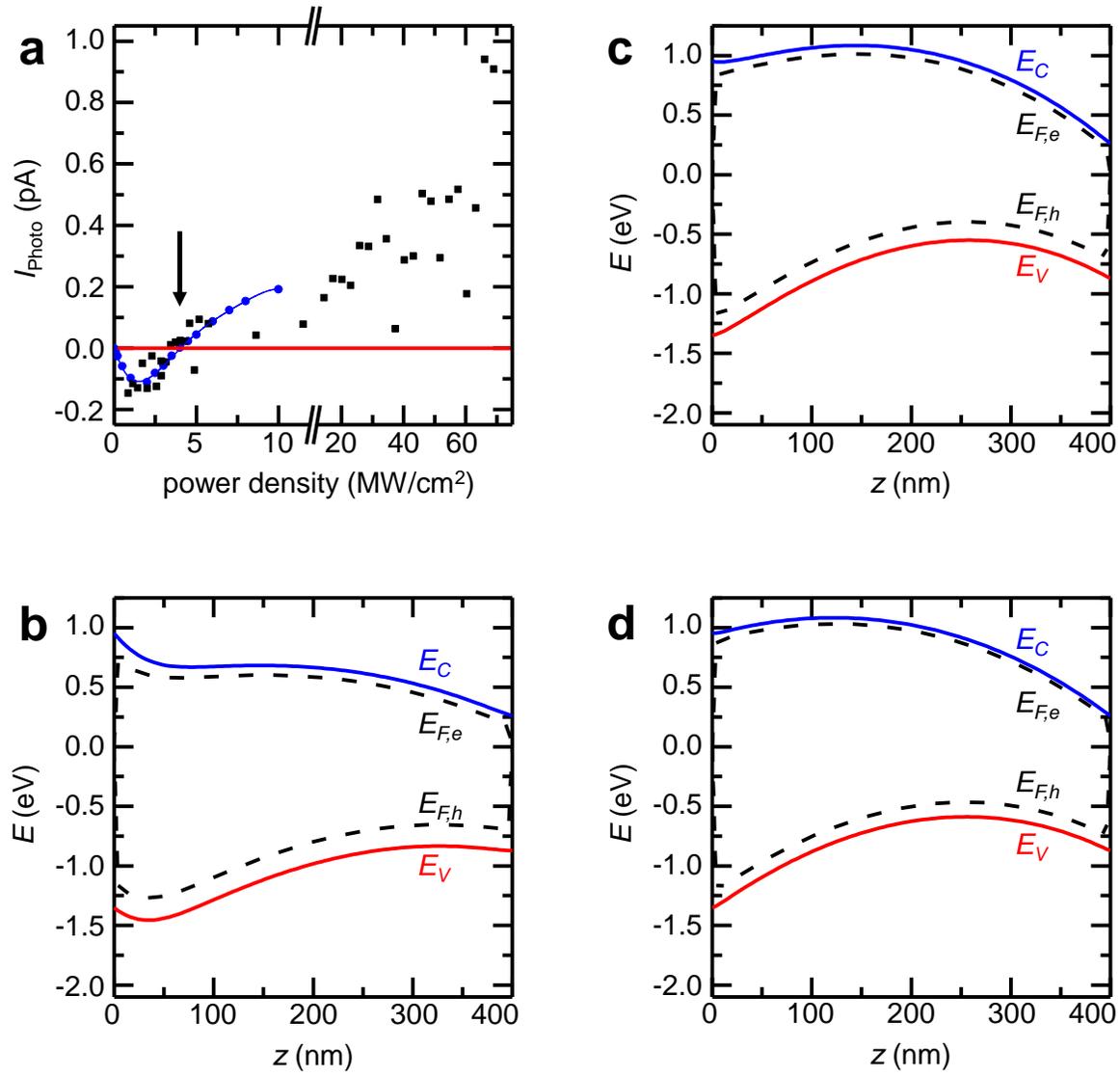

**Figure 4.** (a) Photocurrent $I_{\text{Photo}}$ versus power density for $V_{SD}$ = 0 V ($E_{\text{photon}}$ = 3.1 eV, $T_{\text{bath}}$ = 77 K). Measured data is depicted in black, simulated data in blue. The simulated data is normalized by a proportionality factor to the measurement as the EBID contacts have a considerable resistance reducing the total photocurrent flow. Red line marks the level where $I_{\text{Photo}}$ = 0 A. Band diagram under illumination with a laser power density of (b) 2 MW/cm², (c) 4 MW/cm² and (d) 10 MW/cm².

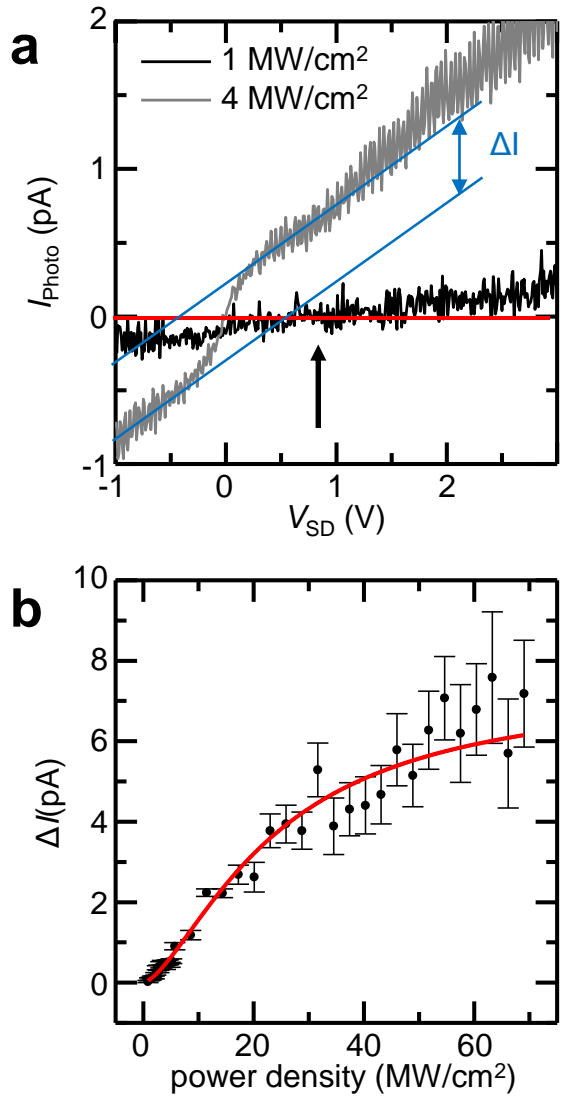

**Figure 5.** (a) Photocurrent-voltage characteristic for 1 MW/cm² (black) and 4 MW/cm² (gray). The black arrow indicates the intersection of zero photocurrent for 1 MW/cm². The red line depicts $I_{\text{Photo}}$ = 0 A. The blue lines define the parameter $\Delta I$ for 4 MW/cm². (b) $\Delta I$ versus power density. The data points are fitted by a saturation function (red line) ($E_{\text{photon}}$ = 3.1 eV, $T_{\text{bath}}$ = 77K).

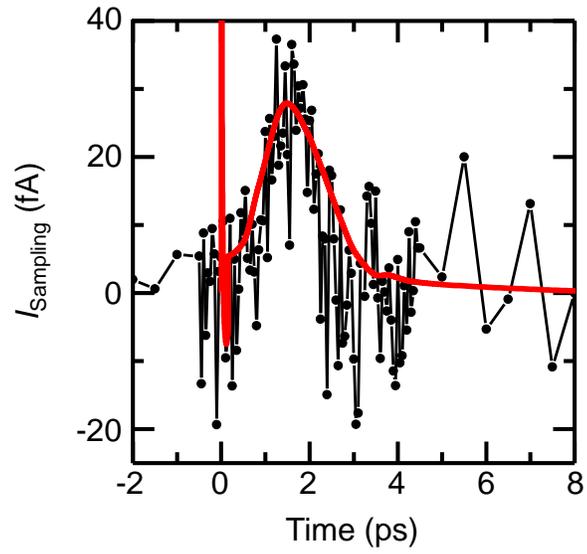

**Figure 6.** Time-resolved photocurrent $I_{\text{Sampling}}$ of the InGaN nanowire. Black dots are the average of 6 single measurements. Red line is the simulated photocurrent ($T_{\text{bath}}$ = 77 K, $V_{\text{SD}}$ = 1 V, $E_{\text{photon}}$ = 3.1 eV, $p_{\text{pump}}$ = 115 MW/cm$^2$, $P_{\text{probe}}$ = 50 mW).

**S1: Nanowire growth parameters.** Catalyst free, self-assembled graded composition InGaN nanowires are grown on a Si(111) substrate by plasma-assisted molecular beam epitaxy (PAMBE) using a Veeco 930 system. The native oxide is removed by heating the substrate at 1100 °C in situ after solvent cleaning. GaN nanowires are nucleated and grown at 720 °C for 20 minutes. The substrate temperature is then reduced to 375 °C. Graded InGaN nanowires are grown on already nucleated GaN nanowire templates using a shutter pulsing method. The shutter open time of the Ga and In cells are altered at each monolayer calculated from previously calibrated GaN and InN vertical growth rates in nanowire form at same conditions. More detail of the growth process can be found in ref. 21 of the main manuscript.

**S2 Time-integrated photocurrent measurement.** The optical excitation occurs by focusing the light of a mode-locked titanium:sapphire laser with a repetition rate of 76 MHz and a pulse duration of 160 fs onto a beta barium borate (BBO) crystal to generate light with 3.10 eV by second harmonic generation. A short pass filter after the BBO crystal blocks photons with energies below 2.21 eV. The frequency doubled light is focused by an objective of a microscope onto the nanowire. The laser spot has an elliptical shape with a major (minor) axis of 2 µm (5 µm) due to second harmonic generation in the BBO. The laser is chopped at a frequency of about 2.3 kHz resulting in a photocurrent $I_{\text{Photo}} = I_{\text{on}} - I_{\text{off}}$ detected with a lock-in amplifier utilizing the reference signal provided by the chopper. $I_{\text{on}}$ ($I_{\text{off}}$) is the current through the nanowire device when the laser is switched on (off).[1]

**S3 Numerical simulation.** Numerical simulations are performed using Technology Computer Aided Design software Silvaco ATLAS. Spontaneous polarization is included in the model. It is important to understand that unlike AlGaN material system, spontaneous polarization charge is weak in InGaN material system. Previously we have shown that our MBE grown GaN nanowires preferably grow in the N-face crystallographic direction[2]. The sample under study has been grown with GaN nanowire nucleation, thus N-face. Considering the grading length of 1 µm and N-face polarity of the nanowires, the negative spontaneous polarization charge can be calculated using

$$\rho_{sp} = \frac{Psp_{InN} - Psp_{GaN}}{L_{grad}} = -8000 \frac{C}{m^2} = -5 \times 10^{16}\ cm^{-3}\ ,$$

where $Psp_{InN}$ (-0.042 C/m²), $Psp_{GaN}$ (-0.034 /m²), and $L_{grad}$ (1 µm) are spontaneous polarization of InN, spontaneous polarization of GaN and grading length, respectively. The negative polarization charge expected to induce p-type conductivity in nanowires with $5 \times 10^{16}\ cm^{-3}$ holes. But it is well known that MBE grown III-N material system suffers from high background n-type doping ($\geq 10^{17}\ cm^{-3}$), specially for samples grown at low temperature (which is the case here). Our energy band calculation considered the effect of spontaneous polarization and background doping ($10^{17}\ cm^{-3}$). The net effect makes the nanowire n-type with $5 \times 10^{16}\ cm^{-3}$ electrons.

For planar InGaN materials system, like GaN/InGaN quantum wells used in planar LEDs, strain induced piezoelectric charge is important and creates strong energy band bending in quantum wells. However, in our structure there is no abrupt heterojunction and the In composition is graded over 1 µm which corresponds to 0.011% lattice mismatch for every nm of composition grading. Moreover, due to high surface to volume ratio, nanowire geometry can effectively relax strain using the side walls. Taking into account of both large grading length (small composition gradient) and nanowire geometry we have ignored any contribution of strain induced polarization charge.

To justify ignoring the strain, we point out that in earlier work, we have carried out full strain modeling[3] in graded InGaN nanowires. In Fig. 4[3], the 3D strain modeling reveals that an InGaN nanowire graded from 0 to 65% In has a non-uniform strain distribution (strain relaxation at surface and along length of nanowire). However, unlike in abrupt junctions which exhibit c-axis (polar axis) strains of more than 1%, the graded structures exhibit strain along the c-direction (polar axis) of <|0.2|%. Those simulations are for a 50 nm graded section. In the current case, we have a 1000 nm long nanowire, which decreases the average c-axis tensile strain to on the order of |$10^{-3}$|% or less. Given the piezoelectric coefficient in InGaN, this level of strain provide at most $5 \times 10^{14}$ cm$^{-3}$ of polarization charge, and as such has negligible impact on the band diagrams.

The Pt workfunction is considered to be 5.6 eV. 160 fs pulses of a 400 nm ($E_{photon}$ = 3.1 eV) monochromatic uniform laser beam is used as a modeled excitation source. Material parameters used in this simulation are taken from ref. 28 of the main manuscript.

Key Statements and models used in ATLAS simulation: beam, mobility, Fermi, srh, incomplete, and fldmob

**S4 In content in the InGaN nanowire.** The as-grown InGaN nanowires have a length of 1 μm. Along the nanowire axis the In content is increased from GaN (nanowire base) to InN (nanowire tip). Due to mechanical transfer onto the sapphire substrate the investigated nanowire has a length of about 400 nm. The In content in the nanowire fragment is investigated by measuring the photocurrent dependence on the energy of the exciting laser. For energies below 1.8 eV, we used the laser emission from a titanium:sapphire laser. Higher laser energies (1.8 eV < E < 2.4 eV) are achieved by guiding the laser light through a nonlinear optical fiber which generates a super-continuum in the visible spectrum. Using band pass filters with a spectral width of about 35 meV, the laser energy is selected. For laser energies above 2.4 eV, we focus the laser light on a beta barium borate (BBO) crystal for second harmonic generation.

The resulting photocurrent is shown in the supporting Figure S1. From a linear fit through the data we get $I_{Photo}$ = 0 pA for a laser energy of (1.13 ± 0.13) eV which we phenomenologically interpret to be the lowest band gap present in the linearly graded InGaN nanowire. A band gap of 1.13 eV is consistent with a maximum In content of (74 ± 2) % in the nanowire fragment.[4]

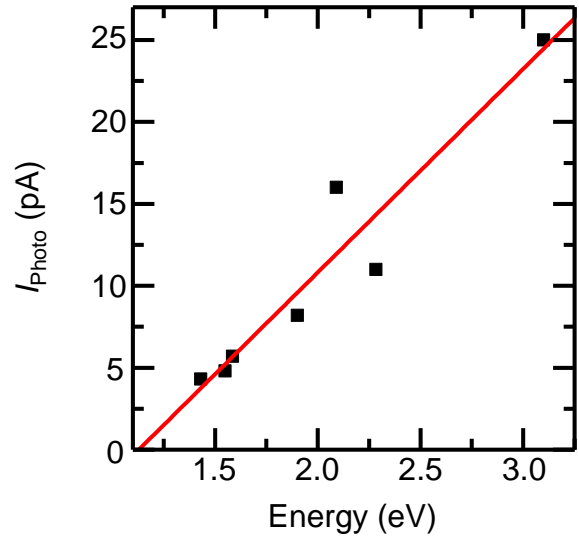

**Figure S1:** Photocurrent dependence on excitation energy. The red line is a linear fit through the measured data points (black squares) ($T_{bath}$ = 77 K, $V_{SD}$ = 1 V, $P_{pump}$ = 60 MW/cm$^2$).

**S2: Picosecond time-resolved photocurrent measurement.** The time-resolved photocurrent $I_{Sampling}$ is measured by the circuitry sketched in the supplementary Figure 2. The InGaN nanowire in the stripline circuit is excited by a pump pulse at a wavelength of 400 nm with an elliptical laser spot with a major (minor) axis of 2 µm (5 µm) and a bandwidth limited pulse duration of about 160 fs. After excitation, an electro-magnetic pulse directly proportional to the photocurrent in the nanowire travels along the strip-line to the field probe where a so-called Auston-switch is excited by a probe laser pulse (800nm, 160 fs) after a time delay Δ*t* controlled by a delay stage. The Auston-switch is made of ion-implanted silicon with a sub-picosecond response time.[5,6] If at the time delay an electro-magnetic transient is present at the Auston-switch, the excited charge carriers by the probe laser pulse amount to current $I_{Sampling}$ in the field probe. Hereby, the photocurrent in the InGaN nanowire is sampled as a function of Δ*t* with a picosecond time-resolution which yields information on the optoelectronic processes in the nanowire.[1,7,8]

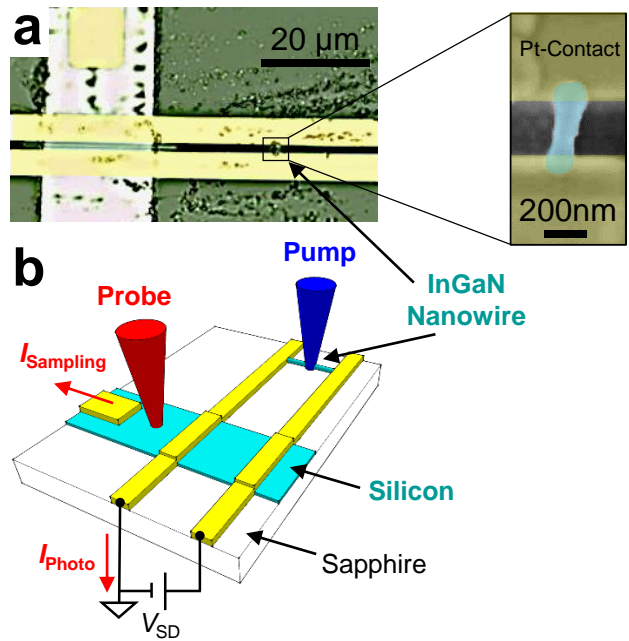

**Figure S2:** (a) Optical microscope image of the sample structure for time-resolved photocurrent measurements with a scanning microscope image of the incorporated InGaN nanowire. (b) Schematic on-chip detection geometry. The pump laser pulse (blue, 400 nm) excites the InGaN nanowire contacted by the gold strip-lines via Pt contacts. The probe laser pulse (red, 800 nm) triggers the photocurrent signal $I_{Sampling}$ which is measured at the field probe. The nanowire and the ion implanted silicon is depicted in turquois, the gold electrodes in yellow.

# Extra References